# Electronic Structure of CeFeAsO$_{1-x}$F$_x$ (x=0, 0.11/x=0.12) compounds

F. Bondino[1], E. Magnano[1], C.H. Booth[2], F. Offi[3], G. Panaccione[1], M. Malvestuto[4], G. Paolicelli[5], L. Simonelli[6,] F. Parmigiani,[1,7] M. A. McGuire,[8] A. S. Sefat,[8] B. C. Sales,[8] R. Jin,[8,9] P. Vilmercati[9], D. Mandrus[8], D.J. Singh[8] and N. Mannella[9,*]

[1] *IOM-CNR, Laboratorio TASC, Basovizza-Trieste, S.S. 14 km 163.5, I-34149 Italy*
[2] *Chemical Sciences Division, Lawrence Berkeley National Laboratory, Berkeley, California 94720, USA*
[3] *CNISM and Dipartimento di Fisica, Università Roma Tre, Via della Vasca Navale 84, I-00146 Roma*
[4] *Sincrotrone Trieste S.C.p.A Area Science Park-Basovizza,, S.S. 14 km 163.5, I-34149 Trieste, Italy*
[5] *CNR-INFM, National Research Center S3, Via Campi 213/aI-41100 Modena, Italy*
[6] *European Synchrotron Radiation Facility, B.P. 220, 38042 Grenoble, France*
[7] *Dipartimento di Fisica, Università degli Studi di Trieste, Trieste, I-34124 Italy*
[8] *Materials Science and Technology Division, Oak Ridge National Laboratory, Oak Ridge, Tennessee 37831, USA*
[9] *Department of Physics and Astronomy, University of Tennessee, Knoxville, TN*
\* *Corresponding author, nmannell@utk.edu*

**Abstract**

We report an extensive study on the intrinsic bulk electronic structure of the high-temperature superconductor CeFeAsO$_{0.89}$F$_{0.11}$ and its parent compound CeFeAsO by soft and hard x-ray photoemission, x-ray absorption and soft-x-ray emission spectroscopies. The complementary surface/bulk probing depth, and the elemental and chemical sensitivity of these techniques allows resolving the intrinsic electronic structure of each element and correlating it with the local structure, which has been probed by extended-x-ray absorption fine structure spectroscopy. The measurements indicate a predominant $4f^1$ (i.e. Ce$^{3+}$) initial state configuration for Cerium and an effective valence-band-to-4$f$ charge-transfer screening of the core hole. The spectra also reveal the presence of a small Ce $f^0$ initial state configuration, which we assign to the occurrence of an intermediate valence state. The data reveal a reasonably good agreement with the partial density of states as obtained in standard density functional calculations over a large energy range. Implications for the electronic structure of these materials are discussed.

**Introduction**

The discovery of high-temperature superconductivity in the Fe-based compounds (FeSC) offers the possibility of studying the relationship between high temperature superconductivity and magnetism in a very large class of materials. Despite the large body of theoretical and experimental work to date, the question of the extent to which the essential physics of the FeSC is similar to that of the cuprates remains controversial. The emergence of superconductivity in close proximity to a long-range-ordered antiferromagnetic ground state and the similarity of the phase diagrams initially suggested a close resemblance between the FeSC and the cuprates [1,2,3,4]. On the other hand, experimental results show remarkable differences between the FeSC and the cuprates. For example, (Ba,K)Fe$_2$As$_2$ has recently been reported to exhibit superconducting properties that are rather isotropic, a behavior drastically different from that of layered cuprates [5]. Angular Resolved Photoemission (ARPES) has directly established the three-dimensional nature of the Fermi surface topology, which has been proposed to explain the relatively weak anisotropy of the critical field of (Ba,K)Fe$_2$As$_2$, thus providing another distinctive difference with the cuprates [5,6,7].

The determination of the most appropriate starting point for a theoretical description of the FeSC remains among the most important unresolved problems in this field. Different theoretical studies have reached opposite conclusions regarding the magnitude of the Hubbard, on-site Coulomb repulsion parameter U. Values of U have been found to range from U = 4 eV [8,9], to $2.2 \leq U \leq 3.3$ [10,11,12,13], to U ≤ 2 [14], thus identifying the FeSC as either moderately strongly correlated or weakly correlated materials, respectively. Particularly relevant in this regard are the results of x-ray spectroscopy experiments, since they are expected to elucidate the role of electron correlations. Indeed, different screening channels of the core hole created upon photon absorption leave definite signatures in photoemission, absorption and x-ray emission spectra typically in the form of additional satellite peaks when correlation effects are at work. These additional spectral features have not been observed in core level photoemission and x-ray absorption experiments in oxypnictide 1111 materials. X-ray spectroscopy experiments have shown that the electronic structure of the normal state is quite different from that of cuprates [15,16], with Fe spectra being characterized by lineshapes more akin to those of Fe metal and intermetallic compounds, in stark contrast to the case of correlated oxides. These results are more in agreement with the existence of weak electronic correlations, with the spectral shapes often being a good match to the occupied/unoccupied electronic density of states determined from standard density functional theory calculations [14]. More recently, x-ray absorption and x-ray emission spectra for different pnictides families have been directly compared to theoretical calculations that included the presence of the core hole in the absorption and emission processes in full detail, with the results indicating that the FeSCs are weakly correlated materials [14]. Other ARPES investigations have identified features of the electronic structures, which are found to be either in agreement [6,17,18], or incompatible with the results of DFT calculations [19,20]. The determination of electronic structure, electronic correlations and to what extent the latter can be linked to the local crystal structure are of fundamental importance for understanding of the macroscopic properties of the FeSC.

In order to provide a comprehensive picture of the electronic structure of FeSC materials, we have studied the electronic and local crystal structure in the normal state of $CeFeAsO_{0.89}F_{0.11}$ and its parent compound CeFeAsO using several x-ray spectroscopy techniques. The F-doped CeFeAsO compounds can be considered representatives of the large group of layered quaternary phases, the so-called oxypnictides, which are formed by alternate LnO (Ln = lanthanide element) and FeAs layers. While the prototypical parent compound CeFeAsO is a non-superconducting metal with a structural distortion below $\approx$ 150 K and long-range spin-density-wave-type antiferromagnetic order, doping the system with F suppresses both the magnetic order and the structural distortion in favor of superconductivity. We have employed techniques with different surface and bulk probing depth, elemental and chemical selectivity, and local structure sensitivity, such as soft and hard x-ray photoemission (SXPES and HAXPES), x-ray absorption (XAS) in total electron yield (TEY) and total fluorescence yield (TFY), x-ray emission (XES) and extended x-ray-absorption fine structure (EXAFS).

The total and partial densities of states (DOS) of the valence band extracted from our data have been compared with Density Functional Theory (DFT) calculations. The photoemission (PES) and XES data, and DFT calculations show consistently that from 0 to 2 eV the total DOS is dominated by Fe $3d$ spectral weight, while Ce $4f$ states are localized around 1.7 eV. The Ce PES core levels spectra indicate a main $4f^1$ initial state configuration and an effective valence-band-to-$4f$ charge-transfer screening of the core hole. Both soft and hard XAS data respectively collected at the Ce $M_{45}$ and Ce $L_3$ edge also indicate that Ce is very close to trivalent, as might be expected. Both XAS and PES spectra show signatures of Fe $d$-electron itinerancy. The spectra do not show large changes in Fe valence with F content. This is an indication of screening, since the Fe bands are being filled with F doping. Furthermore, the data indicate substantial overlap between Fe $3d$ and As $4p$ states and between O $2p$ and Ce $5d$ states, in agreement with the calculations. The good agreement between DFT calculations and experimental data indicate that the correlation effects may not play an important role in these compounds on the energy scale probed by the experiments.

**Methods**

CeFeAsO and $CeFeAsO_{1-x}F_x$ (x=0.11 and 0.12) samples were synthesized using standard solid state techniques [21, 22] from CeAs, Fe, $Fe_2O_3$, and $CeF_3$ mixed in stoichiometric ratios to give the nominal compositions $CeFeAsO_{1-x}F_x$. Powder x-ray diffraction analysis showed the samples to be at least 90% pure, and give lattice constants (*a* and *c*) of 4.00 Å and 8.65 Å, 3.99 Å and 8.63 Å, 3.99 Å and 8.62 Å for x = 0, 0.11, and 0.12, respectively. Magnetization and resistivity measurements revealed superconducting transition onset temperatures near 42 K for x = 0.12 and 38 K for x = 0.11. For x = 0.12, these measurements show clear bulk superconductivity with zero resistivity below 38 K and full diamagnetic screening at low temperatures. Although the transition temperatures are similar, the superconductivity in the x = 0.11 sample appears less robust. The transition is clearly seen both field-cooled and zero-field-cooled magnetic susceptibility measurements, but is partially obscured by a ferromagnetic impurity (probably iron), and the resistivity does not reach zero at low temperatures. Powder diffraction patterns do not suggest significantly increased inhomogeneity in the x = 0.11 sample, so impurity phases at grain boundaries may be partially responsible for the observed behavior.

The soft x-ray photoemission (SXPES), x-ray emission (XES) and x-ray absorption (XAS) measurements (40 eV<hν<1600 eV) were carried out on the BACH beamline at the Elettra Synchrotron Facility [23]. The XAS data have been obtained both in total electron yield (TEY) by measuring the drain current, and in total fluorescence yield (TFY) with a silicon photodiode with the sample biased at a positive voltage. To ensure reproducibility of the data, several samples have been measured at room temperature in a pressure better than $4\times10^{-10}$ mbar after being fractured or repeatedly scraped *in-situ* using a diamond file. The hard x-ray photoemission (HAXPES) measurements (hν=7596 eV) were carried out on beamline ID16 at the ESRF Synchrotron Facility using the Volume Photoemission (VOLPE) spectrometer [24]. The HAXPES measurements were performed with the photon beam impinging at 45° onto the sample surface and by collecting photoelectrons at normal emission, in a pressure better than $2\times10^{-9}$ mbar, from samples whose surface was repeatedly scraped in situ using a diamond file. The Fermi level and the energy resolution were determined from the Fermi level of a gold sample. All measurements have been carried out at room temperature. The experimental energy resolutions of the photoemission spectra has been set to 0.08 eV, 0.19 eV, 0.34 eV, 0.34 eV, 1 eV and 0.47 eV at excitation energies hν=175 eV, 456.4 eV, 882 eV, 700 eV, 946 eV and 7596 eV, respectively. The experimental resolution of the XAS spectra has been set at 0.15 eV at the O $K$ edge, 0.34 eV at the Ce $M_5$ edge, 0.20 eV at the Fe $L_3$ edge.

Ce $L_3$-, Fe $K$-, and As $K$-edge x-ray absorption near edge structure (XANES) data were collected on beamline 11-2 at the Stanford Synchrotron Radiation Light source using a half-tuned Si(220) double-crystal monochromator with a $\phi = 90°$ face. Harmonic content in the beam was further reduced in the Ce and Fe edge data using a harmonic-rejection mirror with a 10.5 keV cut-off energy. Fe $K$-edge data were also collected on beamline 4-1 with a similar monochromator, with $\phi=0°$ face, but with no rejection mirror. Samples were ground under acetone, passed through a 20 μm sieve, brushed onto adhesive tape, and stacked to achieve a change in the absorption across an edge $\Delta\mu t \sim 1$. The samples were placed in a liquid-helium flow-cryostat, and data were collected between 20 K and 300 K. Extended x-ray absorption fine structure (EXAFS) measurements have been carried out at the As and Fe $K$ edges. Fits to the local structure around the Fe and As atoms followed standard procedures [25]. As $K$-edge data were collected at temperature from 20-300 K, while Fe $K$ edge data were collected from 50 K to 300 K.

Density functional calculations were performed within the local spin density approximation (LSDA) for CeFeAsO and for 10% F doped CeFeAs(O,F) using the experimental structures and atomic positions. The doping was treated within the virtual crystal approximation on the oxygen site as in our prior studies [26,27]. For this purpose we used the general potential linearized augmented planewave method as implemented in the WIEN2K package [28] with tested basis sets and zone samplings. The final zone sampling used 440 **k**-points in the symmetry irreducible wedge of the zone. The LAPW sphere radii were 2.35 Bohr for Ce, 2.2 Bohr for Fe and As and 1.8 Bohr for O. Local orbitals were used to relax the linearization of the Fe d bands and to treat the semi-core states. [29] The Ce f-orbitals were treated using the standard LDA+U method, with $U_{eff}$=7 eV and a nearest neighbor antiferromagnetic ordering on the Ce sub-lattice. We emphasize that no U was applied to Fe. The projections of the electronic densities of states were obtained by integration within the LAPW spheres.

**Results and Discussion**

The local order of the samples have been investigated by bulk-sensitive EXAFS measurements of the As and Fe $K$ edges (Fig. 1). All compounds are structurally very well ordered, with no anomalies occurring either with temperature, transition-metal species, or fluorine concentration. The detailed results, not reported here for the sake of brevity, do not indicate any observable departures from the long-range structure inferred from diffraction results [21], although some details bear emphasizing. First, As *K*-edge data were collected at temperature from 20 to 300 K. The fits to these data are excellent [Fig. 1 (a)], with an overall amplitude reduction factor $S_0^2=1.00\pm0.05$, bond lengths consistent with diffraction measurements, and mean-squared correlated displacement (i.e. Debye-Waller) factors, $\sigma^2$, that increase with temperature consistent with a correlated-Debye model [30] and no measurable disorder. Fe *K*-edge data were collected from 50 K to 300 K, with an overall amplitude reduction factor $S_0^2=0.80\pm0.05$, with similar data quality and fit results [Fig. 1 (b)]. These results indicate the samples are locally ordered around Fe and As atoms.

Fig. 2 shows the As $2p_{3/2}$, Ce 4*s*, C 1*s*, O 1*s* core level and shallow core level HAXPES spectra obtained from $CeFeAsO_{0.89}F_{0.11}$ before and after scraping the sample in UHV with a diamond file. Following experimental determinations [31] and theoretical calculations [32], we estimate that the HAXPES measurements here reported, obtained with excitation energy of 7596 eV at normal-incidence (having kinetic energy $E_k$: 6230 eV<$E_k$<7592 eV), have an effective attenuation length (EAL) of about 70 Å. The corresponding probing depth (defined as the overlayer thickness from which 95% of the total signal is produced) is of the order of 200 Å. This value is much higher than the typical EAL of SXPES, which is about 5-20 Å in a kinetic energy range of 20-1400 eV [31,32]. In spite of the relatively bulk sensitive conditions of the experiment, clear signatures of surface contamination, indicated with arrows in Fig. 2, are still detected in the sample before scraping in UHV. The surface contamination components appear more clearly in the As 2*p* [Fig. 2 (b)], As 3*p* and As 3*d* [Fig. 2 (a)] core level spectra as new peaks located at higher binding energy (BE) originating from surface arsenide oxides [33]. Furthermore, some spectral weight is also visible at the high BE side of Ce 4*d* core level [Fig. 2 (a)], likely originated from a surface cerium oxide component having a Ce $4f^0$ ground-state configuration, as discussed more extensively below. In the clean sample, the C 1*s* peak close to Ce 4*s* core level is completely suppressed [Fig. 2 (c)], and the O 1*s* line appears as a single component with strong line asymmetry [Fig. 2 (d)].

An overview of the shallow core level photoemission spectra of $CeFeAsO_{0.89}F_{0.11}$ obtained with soft (456.4 eV) and hard-x-ray (7596 eV) excitation energies is shown in Fig. 3. The marked changes in the relative intensity of the peaks are due to the cross-section variation of electrons with different orbital character excited by soft and hard x-rays. In particular, with increasing photon energy the cross sections of *s* and *p* states decrease much less rapidly than the cross sections of *d* and *f* states [34,35,36]. This marked photon energy dependence of the photoelectron cross section is very useful for identifying the shallow core levels and disentangling the *s/p* and *d/f* contributions in the valence band spectra. In particular, as the photon energy increases, the As 3*d* core level is strongly suppressed, while the peak at 37 eV, identified with Ce 5*s* [37], is well visible. A peak at 12 eV is clearly visible in the HAXPES spectrum, while in the SXPES spectrum is barely visible. A similar peak was also observed in other FeAs compounds [38]. The photon energy dependence of the cross-section indicates that this peak has likely *s* or *p* orbital character.

A comparison of valence band HAXPES, SXPES, and XES spectra extending from the Fermi level ($E_F$) to 15 eV below $E_F$ is shown in Fig. 4 for $CeFeAsO_{0.89}F_{0.11}$. This is compared with the results of DFT calculations for 10% F doping. Since our samples are polycrystalline, the measured angle-integrated photoemission valence band spectra provide a representation of the occupied DOS weighted by the orbital cross section. A strong suppression of the spectral weight at the $E_F$ occurs with increasing the photon energy from 175 eV to 7596 eV, suggesting that the states at $E_F$ have a predominant *d/f* character, which is much stronger in SXPES than HAXPES, as discussed above. This assessment is supported by the partial DOS calculated with DFT [cf. Fig. 4 (d-e)]. The calculations indicate that the states close to $E_F$, which are enhanced in the SXPES spectrum, are dominated by the Fe 3*d* DOS, in agreement with the conclusions derived from our previous study [15]. The O 2*p* and As 4*p* states are placed at ≈ 3 and 5 eV below $E_F$, in correspondence to the main structure in the VB HAXPES spectrum. The Ce 4*f* states are located in close proximity to the value 1.7 eV, as obtained from resonant SXPES measurements across the Ce $M_5$ edge [15]. The calculations reveal overlap between O 2*p* and Ce 5*d* states at ≈ 5 eV and between As 4*p* and Fe 3*d* states at ≈ 3 eV. The calculations identify the peak at ≈ 12 eV with the As 4*s* shallow core level, consistent with the fact that its intensity is markedly enhanced in the HAXPES spectrum. This assignment is different from that proposed by Ding et al. [38] in $Ba_{0.6}K_{0.4}Fe_2As_2$, namely a satellite state with an Fe $3d^5$ configuration suggesting the importance of local electronic correlations at the Fe sites. On the contrary, the agreement between our data and the DOS determined from DFT calculations indicates weak electronic correlations [14].

Fig. 4 (c) shows the F $K_\alpha$, O $K_\alpha$, Ce $M_\alpha$ and Fe $L_\alpha$ XES spectra measured at room temperature. The spectra are aligned to a common energy scale obtained by subtracting the BE of the excited core level from the emission photon energies calibrated using elastic peaks as references, a procedure which allows obtaining an experimental determination of the partial occupied DOS. Considering the dipole selection rules and the major contribution expected from some transitions (e.g. the transition probability of $3d \rightarrow 2p$ being much stronger than the one of $4s \rightarrow 2p$), the F $K_\alpha$, O $K_\alpha$, Ce $M_\alpha$ and Fe $L_\alpha$ XES spectra can be associated to F $2p$, O $2p$, Ce $4f$ and Fe $3d$ bulk DOS, respectively. The position of Ce $4f$ states is consistent with the LDA+U calculations and our previous finding obtained from Ce $M_5$ resonant SXPES [15]. The agreement between experimental and theoretical partial DOS is extremely good, suggesting that DFT approach is appropriate to describe the electronic structure of these compounds.

We specify that the F $K_\alpha$, O $K_\alpha$, Ce $M_\alpha$ and Fe $L_\alpha$ XES spectra have been measured with excitation energies of 696 eV, 534.1 eV, 882.4 eV, 717 eV, respectively. These excitation energies are high enough to be in the normal fluorescence regime for all of the spectra, and lower than the onset energy for the appearance of high-energy satellites due to double and triple vacancy configurations, which can appear in XES spectra as spurious peaks having no correspondence with partial DOS [39,40,41]. For this reason, caution has to be exercised when interpreting F $K$ and O $K$ XES spectra as partial DOS when spectra are obtained with excitation energy above the onset for $K^{-1}L^{-1}$ hole excitation. A compelling example of this state of affairs is provided by O $K_\alpha$ XES spectra excited with two different excitation energies, as shown in Fig. 5. The excitation energy of 731 eV gives rise to the appearance of a high-energy shoulder due to spectator holes in the $L$ shell. The high-energy shoulder is instead absent in the spectrum measured with an excitation energy of 534.1 eV, below the onset for $K^{-1}L^{-1}$ hole excitation.

Fig. 6 shows a series of $L_\alpha$ XES spectra obtained at resonant excitation energies set across the Fe $L_{23}$ XAS edge [Fig. 6 (a)]. The lineshape of Fe $L_{23}$ XAS does not exhibit clear multiplet satellites. Rather, it is typical of systems with a high degree of electron delocalization. The weak and broad shoulder at 709.5 eV is also present in the Fe XAS spectra of Fe-X (X is an $sp$ element) compounds with strong transition metal $3d$-X $np$ hybridization such as FeSi [42], MnSi [43] and $Mn_5Ge_3$ [44]. This shoulder is thus indicative of a covalent nature of the As and Fe conduction electrons in the FeAs plane in the sense that As provides screening. This is also reflected in the DFT calculations (Fig. 4). In the series of Fe $L_\alpha$ XES spectra no inelastic structure associated to low-energy excitations are detected, at variance with what is found in Fe oxides. The resonant XES spectra display only structures at constant emission energy, at 705 eV and 717 eV. This behavior is typical of metallic systems such as Fe metal, consistent with an absence of strong correlations in this system. We notice that similar results have been obtained in several other 1111 or 122 FeAs superconductors [14].

We now turn on the discussion of As and Ce photoemission and XAS spectra. The As $3d$ core level PES spectra measured with two different photon energies (456.4 eV and at 7596 eV) are shown in Fig. 7. It is clear that more than one component is needed in order to fit the data. For instance, in the bulk-sensitive HAXPES measurement two main peaks are visible, but their branching ratio and the spin-orbit splitting clearly indicates that they cannot be attributed to the spin-orbit doublet (As $3d_{5/2}$ and As $3d_{3/2}$), but they rather originate from two different bulk As species. We have fitted the spectra according to a procedure consisting of a removal of an integrated background and fit with a Doniach-Sunjic lineshape convoluted with a Gaussian function of different width for each spectral component (e.g. different width is expected for surface components or when disorder or non-equivalent sites are present). In all of the fits the atomic parameters such as the Lorentzian Full width Half Maximum (FWHM) and the spin-orbit splitting were kept constant at 0.16 eV and 0.69 eV, respectively [33]. The branching ratio was allowed to vary ±10% around the statistical value since this parameter can show some degree of anisotropy, depending both on the structure and the photon energy [45]. The results of this fitting procedure are shown as continuous lines in Fig. 7 along with the experimental data. The more bulk-sensitive spectrum measured at 7596 eV can be fitted using two components denoted as B1 and B2. We find that six components, denoted as B1, B2, S, O1, O2 and O3, are needed to fit the more surface sensitive spectra measured with 456.4 eV. The intensity of the broad peaks O1, O2 and O3 decreases with increasing photon energy, indicating that they can be attributed to surface components, likely originating from surface arsenic oxides as the result of a small residual surface contamination observed in the more surface sensitive SXPES spectra. Indeed, these components have also been observed in the bulk-sensitive HAXPES spectra before scraping in UHV [cf. Fig. 2 (a)]. The S and B1 components are much narrower than B2, O1, O2 and O3. The S component is absent in the HAXPES spectrum, and its intensity decreases with increasing photon energy, suggesting that it can be attributed to a surface component. We notice that two components with the same BE as those of the S and B1 components have been observed in $BaFe_2As_2$ by de Jong et al. [46] and attributed to a surface and bulk components, respectively, in agreement with our conclusions. The similarity in BE position and linewidth of the S and B1 components to those observed in the $BaFe_2As_2$ system suggests that they can be assigned to the As atoms inside the As-Fe plane. Compared to $BaFe_2As_2$, $CeFeAsO_{0.89}F_{0.11}$ exhibits an additional component B2 with larger linewidth. Since this component is visible in the HAXPES spectrum, it is undoubtedly of bulk origin. We tentatively attribute this component B2 to interplanar As-O bonds. This interpretation is consistent with its absence in the As $3d$ spectra measured in oxygen-free compounds such as $BaFe_2As_2$ [46,47] and $Ba_{0.6}K_{0.4}Fe_2As_2$ [38]. This interpretation would also be consistent with the As $3d$ spectra measured by Koitzsch et al. in $LaFeAsO_{1-x}F_x$ compounds [48]. Koitzsch et al. observed a doping dependence of the As $3d$ lineshape, with the highest BE As $3d$ component (corresponding to B2 in our spectra) being most intense in the undoped compound. This observation is fully consistent with our interpretation, since the undoped compound in the series $LaFeAsO_{1-x}F_x$ is the one with highest O content, and hence As-O bonds. Both in HAXPES and SXPES spectra, the B1 component is more intense than the B2, while the B1/B2 intensity ratio is higher in the SXPES spectrum. The reason of this energy dependence is not clear and needs further investigation.

The structures found in the Ce $M_{45}$ XAS spectrum measured from the sample after scraping in UHV (Fig. 8) correspond quite well to the energy of the $3d^94f^2$ configuration, with the intensity distribution being well reproduced by the electric dipole transition from the $J = 5/2$ ($4f^1$) ground state to the $3d^94f^2$ final state [49]. Thus the occupation of the $f$ states is mainly $4f^1$, corresponding to a +3 charge state. The lineshape is consistent with a nearly localized character of the Ce $4f$ electrons [50]. This is also what we find in our LDA+U calculations. We note that additional structures related to $4f^0$ initial-state ($3d^94f^1$ final-state) configuration appear when XAS spectra are collected in the sample exposed to air. The occurrence of these components is consistent with the presence of a more pronounced Ce $4f^0$ spectral weight in Ce $4d$ core level HAXPES spectrum measured in the sample before scraping [cf. Fig. 2 (a)]. These structures are strongly suppressed after scraping, indicating that, if compared to systems like $CeRu_2$, the amount of mixed valency is small in these compounds [50]. This is fully consistent with the bulk sensitive XAS measurements at the Ce $L_3$ edge, shown in Fig. 9 together with a $CeO_2$ spectrum. A direct comparison with the $CeO_2$ spectrum is very valuable for the determination of the Ce valence. As in $CeO_2$, any tetravalent Ce component would manifest as a distinct feature near the second maximum ($\approx 5737$ eV), as well as a large shoulder or distinct feature due to the first maximum at $\approx 5729$ eV [51]. The $CeFeAsO_{1-x}F_x$ (x=0, 0.12) data show a small bump that is possibly due to a $4f^0$ component of Ce, although the feature may be structural in origin. Assuming the feature is only due to a $4f^0$ component, these data indicate about $\approx 5\%$ $4f^0$ Ce component in these compounds. The spectral change in the white-line height from the Ce $L_3$ edge appears to conserve overall area. The changes in height in the main peak of the Ce spectrum are not currently understood, but may relate to electronic differences between the samples, possibly as a result of small impurity distributions or different hybridization of the Ce $4f$ states, an important point that we will address below. Data up to T = 100 K indicate no changes with temperature.

Further insights in the electronic ground state of Ce are provided by photoemission experiments. Ce is an extremely reactive element, and particular care has to be taken in the determination of its ground state configuration. The use of HAXPES has proved very useful in this particular issue because of the bulk sensitive character of the measurement. Fig. 10 shows the Ce $3d$ and Ce $4d$ HAXPES core level spectra measured at room temperature in a clean $CeFeAsO_{1-x}F_x$ (x=0, 0.11) samples, with no detectable traces of surface oxides. One notable feature is the shape of the Ce $3d$ core level peak [Fig. 10 (a)], as it appears that the $3d_{5/2}$ and $3d_{3/2}$ peaks, centered at $\approx 885$ eV and $\approx 903$ eV and separated by an amount equal to the spin orbit splitting ($\approx 18$ eV), are composed of two doublet structures. These doublet structures are well known to arise from two different ways of screening the initially empty 4$f$ orbital upon creation of the 3$d$ core hole that is left behind due to the photoemission process [52]. For the so-called "poorly screened" or "unscreened" channel, the 4$f$ level is predominantly screened by electrons belonging to the outermost shell of the Ce atom, while for the so-called "well-screened" channel electrons belonging to the ligand atom (O in this case) are transferred to the 4$f$ subshell to screen the 3$d$ hole. The two unscreened peaks thus correspond to the 4$f^1$ [Ce$^{4+}$(3$d^9$5$p^6$4$f^1$)*] final state, and the well screened peaks to the 4$f^2$L$^{-1}$ [Ce$^{3+}$(3$d^9$5$p^6$4$f^2$)* + O$^{1-}$(2$p^5$)*] final state, where L$^{-1}$ denotes a hole in the ligand, and the asterisk means that these configurations are excited states. The strong intensity of the $f^2$ peak, which can be qualitatively regarded as the hybridization strength, indicates that the valence-band-to-4$f$ charge-transfer screening of the core hole is very effective in this compound. In addition to the 4$f^1$ and 4$f^2$ components, also visible is a broad peak located at higher BE, and identified as a 4$f^0$ [Ce$^{5+}$(3$d^9$5$p^6$)*] final state configuration. Since the HAXPES measurements are bulk sensitive, the observation of a weak $f^0$ configuration is intrinsic, and not related to surface contamination. The detection of the $f^0$, $f^1$ and $f^2$ configurations in the bulk sensitive HAXPES spectra indicates the presence of a mixed valence ground state of $f^1$ and $f^0$ character. The $f^1$ ground state component (i.e. Ce$^{3+}$) gives raise to well-screened $f^2$ and unscreened $f^1$ final states, while the $f^0$ ground state component (i.e. Ce$^{4+}$) is associated with the well-screened $f^1$ and unscreened $f^0$ final states. The Ce 4$d$ spectra of CeFeAsO$_{1-x}$F$_x$ (x=0, 0.11) [cf. Fig. 10 (b)] are characterized by the presence of these structures as well, but the reduced spin-orbit separation (3.3 eV [53]) forces the screened and unscreened peaks of the Ce 4$d_{5/2}$ and 4$d_{3/2}$ core levels to overlap, resulting in a more complex shape of the overall spectrum. A weak hump is visible at $\approx 125$ eV, probably arising from a minor contribution of the 4$f^0$ ground state configuration. Although being only a minor part of the ground state configuration, the origin of the $f^0$ ground state component may be indicative of important details in the electronic structure, as we now comment below.

The detection of an $f^0$ ground state component (i.e. Ce$^{4+}$) may be associated either with the presence of a small percentage of CeO$_2$ impurities or the occurrence of an intermediate valence state. Although no diffraction peaks associated with either CeO$_2$ or other Ce oxide phases have been observed, the presence of small (i.e. < 5%) impurities cannot be completely excluded due to the detection limits of our instrument. The presence of CeO$_2$ impurities can be consistent with the signatures in the Ce L$_3$ edge XAS [cf. Fig. 9] and Ce 3$d$ HAXPES [cf. Fig. 10 (a)] spectra. It is nonetheless puzzling to observe that the Ce M$_{4,5}$ XAS spectrum [cf. Fig. 8], which has a probing depth lower than that of the Ce 3$d$ HAXPES spectrum, does not exhibit signatures of oxide phases, thus questioning the interpretation of an impurity-related origin of the $f^0$ ground state component.

An intermediate valence state would occur in the presence of some degree of hybridization of Ce 4$f$ states with the valence band electrons, with the energy difference between 3$d^9$4$f^0$ and 3$d^9$4$f^1$ final configurations being small. In CeFeAsO$_{1-x}$F$_x$ there is some degree of Ce 4$f$-O 2$p$ mixing as revealed both by the partial density of states (Fig. 4) and the significant weight of the $f^2$ component in the Ce 3$d$ core level spectra. This mixing can be responsible of a moderate coupling strength. In the presence of a core-hole, if the energy difference between the $f^1$ and $f^0$ configurations is small, this moderate coupling strength would allow for a mixing of the configurations 4$f^0$ and 4$f^1$ in the ground state, which will give rise to three different configurations in the final state, namely 4$f^0$, 4$f^1$, and 4$f^2$. Although from the M$_{45}$ Ce XAS spectra the ground state appears almost pure 4$f^1$, the presence of $f^0$ and $f^2$ components in the Ce 3$d$ spectrum and the small $f^0$ in the Ce L$_3$ XAS spectra are consistent with the presence of a small, but still detectable hybridization and a mixed valence state $f^0$-$f^1$ (i.e. Ce$^{4+}$/ Ce$^{3+}$), with $f^1$ (Ce$^{3+}$) being highly predominant. Importantly, the lattice constants of LnFePnO (Ln=La,Ce,Pr,Nd,Sm,Gd; Pn=P, As) across the Ln series exhibit a deviation at Ce which is consistent with a small amount of mixed valence [22,54]. Most notably, for CeFeAsO a specific heat gamma coefficient of 50 mJ/molK$^2$ has been reported, about an order of magnitude larger than that of LaFeAsO, and attributed to the presence of "pronounced correlations" of the 4$f$ electrons in the ordered state [55]. It is possible that the enhancement of the specific heat coefficient is related to the presence of a mixed valence, even if the analysis of the low T heat capacity is probably complicated by the low Ce ordering temperature (4 K in CeFeAsO). In light of these facts and the absence of significant impurity CeO$_2$ components in the M$_{45}$ Ce XAS spectra, which have a lower probing depth than the Ce 3$d$ HAXPES spectra, we propose that the origin of the $f^0$ ground state component inferred from our measurement is a mixed valence state.

The peak extending on the lower BE side of the Ce 4$d$ core level is the Fe 3$s$. The Fe 3$s$ core level measured with HAXPES (hv = 7596 eV) exhibits a clear multiplet splitting, whose importance related to the presence of spin fluctuations has been discussed in our previous work [15]. As shown in the inset of Fig. 10 (b), the Fe 3$s$ HAXPES core level spectrum exhibit a splitting equal in magnitude to that obtained with SXPES, thus confirming the intrinsic bulk character of the multiplet splitting and consequently the presence of spin fluctuations [15].

To illustrate the effect of F doping, in what follows we compare the electronic structure of the undoped and doped compounds. The bulk-sensitive hard x-ray XANES spectra of Ce $L_3$- (Fig. 9), Fe $K$- [Fig. 11 (a)], and As $K$-edge [Fig. 11 (b)] show no obvious indications of a change of valence with fluorine concentration. As for the Ce $L_3$ edge, the spectral change in the white-line height in the 1$s$-3$d$ feature at 7114 eV in the Fe $K$ edge appear to conserve overall area, and so is more likely due to the presence of a small percentage of impurities.

Figs. 12 and 13 show a comparison of several XAS, PES and RXES spectra of CeFeAsO$_{0.89}$F$_{0.11}$ and CeFeAsO measured at room temperature. The Fe 2$p$ SXPES core level spectra [Fig. 12 (a)] do not display satellite features commonly associated with 3$d$ electrons with localized character, as found for example in the Cu 2$p$ spectra of the high-$T_C$ cuprates superconductors or Fe oxides. Fe 2$p$ PES spectra are more akin to those found in systems with a remarkable degree of electron delocalization such as metallic iron, a fact that is indicative of the absence of strong electron correlation and localization effects. A comparison of the valence band PES spectra from CeFeAsO$_{0.89}$F$_{0.11}$ and CeFeAsO [Figs. 12 (b) and (c)] reveal noticeable differences in the spectral weight located at ≈ 9 eV. Since the spectral weight increases for the doped sample, we assign the structure at 9 eV to the F 2$p$ DOS. This assignment is in agreement with the F 2$p$ DOS obtained experimentally from XES spectra (cf. Fig. 4). The F 2$p$ DOS could not be obtained in the virtual crystal approximation used for the present DFT calculations. No significant differences instead are found between doped and parent compounds in the lineshape and intensity of the states close to the Fermi level (Fe 3$d$ and Ce 4$f$). Although the electron count is increased with F doping, the position of the Fermi energy relative to the Fe core states changes very slightly. For example, the change in the calculated Fe 1s binding energy relative to the Fermi level is less than 10 meV. This is consistent with the non-rigid band behavior upon doping noted for the BaFe$_2$As$_2$ and LiFeAs [56]. No clear spectral differences are detected in the Fe $L_{23}$ XAS and Ce $M_{45}$ XAS spectra in the doped and undoped compound [Figs. 13 (a)-(b)]. Besides the obvious presence of F $K$ emission in CeFeAsO$_{0.89}$F$_{0.11}$, no differences are found also for the Fe $L_\alpha$ XES spectra between doped and parent compounds, as shown in Figs. 13 (c)-(d).

We measured O $K$ edge XAS spectra both in TEY and in the more bulk sensitive TFY mode. The O $K$ edge XAS spectrum involves the transition from the O 1$s$ core level to the empty O 2$p$ states and it can be considered as representative of O 2$p$ unoccupied DOS in the conduction band, possibly hybridized with $s/d$ states of the other elements. The O $K$ XAS lineshape is almost identical in both compounds, as shown in Figs. 13 (e)-(f) for spectra collected in TFY mode. The shape of O $K$ XAS spectra collected in TEY and TFY is qualitatively identical, with only few differences due to possible saturation effects in the TFY spectra. Other groups detected a similar O $K$ XAS lineshape for similar compounds having a different lanthanide element, such as LaFeAsO$_{1-x}$F$_x$ [57]. A zoom of the pre-peak region of the O $K$ edge [Fig. 13 (f)] reveals that within experimental uncertainty (0.15 eV) the O $K$ edge does not display the doping dependence in energy shift observed in ref. [57]. We have not observed energy shifts in the spectra measured in TEY as well. The presence of a pre-peak is quite interesting, as it immediately indicates that the 2$p$ shell of the O atom is not completely full, i.e. O is not O$^{2-}$. Since O is mainly coordinated to Ce, we interpret the presence of the pre-peak as a signature of the hybridization between Ce 5$d$ with O 2$p$ states. This interpretation is supported by a direct comparison of the O $K$ XAS spectrum with the partial DOS DFT calculations for Ce 5$d$ and O 2$p$ states. As shown in Fig. 14, the empty O 2$p$ and Ce 5$d$ states overlap in a large energy region located in correspondence of the pre-peak. The presence of the pre-peak is consistent with the 4$f^2$ [Ce$^{3+}$(3$d^9$ 5$p^6$ 4$f^2$)* + O$^{1-}$(2$p^5$)*] excited configuration detected in Ce 3$d$ and Ce 4$d$ SXPES spectra, as discussed above.

The absence of any noticeable F dependence in PES, XAS and XES spectra as well as in the hard x-ray XANES spectra of Fe $K$- and Ce $L_3$- edges is quite surprising given that F doping induces superconductivity. It is nonetheless consistent with the DFT predictions, namely that the main effect induced by doping is an upward shift of the chemical potential in the Fe d bands, as confirmed for example in the Co doped 122 materials [6]. This is an indication of screening, since the Fe bands are being filled with F doping. However, it should be emphasized that our experiments probe an energy range which extends beyond the energy of the main d bands, which may well be renormalized (as indeed shown by many ARPES investigations) and exhibit changes with doping.

**Concluding remarks**

The electronic and local crystal structure in the normal state of the Fe-based high temperature superconductor $CeFeAsO_{0.89}F_{0.11}$ and its parent compound CeFeAsO has been comprehensively studied with several x-ray spectroscopy techniques such as soft and hard x-ray photoemission spectroscopy (SXPES and HAXPES), x-ray absorption spectroscopy (XAS) in total electron yield (TEY) and total fluorescence yield (TFY), x-ray emission spectroscopy (XES) and extended x-ray-absorption fine structure (EXAFS). Implications for the electronic structure of these materials are discussed on the basis of density functional calculations performed within the local spin density approximation (LSDA).

The different surface and bulk probing depths, elemental and chemical sensitivity have been exploited to provide a comprehensive description of the electronic structure. The bulk-sensitive EXAFS data collected at the Fe and As edges indicate that the compounds are very well ordered, with no anomalies occurring either with temperature, transition-metal species, or fluorine concentration. PES and XES data reveal that within 2 eV below $E_F$ the total DOS is dominated by Fe 3d spectral weight, with XAS and PES spectra showing signatures of Fe d-electron itinerancy. The data indicate the presence of a strong overlap between Fe 3d and As 4p states and between O 2p and Ce 5d states. The occupied Ce 4f states are located at ≈ 1.7 eV below $E_F$. The Ce core level PES spectra indicate a main $4f^1$ initial state configuration and an effective valence-band-to-4f charge-transfer screening of the core hole. The assessment of $4f^1$ initial state configuration, i.e. $Ce^{3+}$, is consistent with soft and hard XAS data collected at the Ce $M_{45}$ and Ce $L_3$ edge, respectively. The spectra also reveal the presence of a small $f^0$ initial state configuration, which we assign to the occurrence of an intermediate valence state. All the spectroscopy data indicate that little change occurs in the Fe valence with fluorine concentration. The total and partial DOS extracted from the experimental data have been found to be in good agreement with those calculated with standard DFT calculations. It thus appears that that despite some known problems of DFT calculations, such as a strong overestimation of the magnetic tendency of these materials and difficulties describing the interplay between magnetism and Fe-As bonding [58], the description of the electronic structure concerning the orbital occupancies and their relative energies in spectra are not strongly modified by electron correlations. These findings are quite different from what is expected in strongly correlated oxides, such as cuprates, and impose stringent constraints on theories capable of providing a correct description of FeSC materials.

**Acknowledgments**

The work at Elettra and European Synchrotron Radiation Facility is supported by NSF grant DMR-0804902. We acknowledge Elettra and the European Synchrotron Radiation Facility for provision of synchrotron radiation facilities. The work at Oak Ridge is sponsored by the Division of Materials Science and Engineering, Office of Basic Energy Sciences. Oak Ridge National Laboratory is managed by UT-Battelle, LLC, for the U.S. Department of Energy under Contract No. DE-AC05-00OR22725. Portions of this research performed by Eugene P. Wigner Fellows at ORNL.

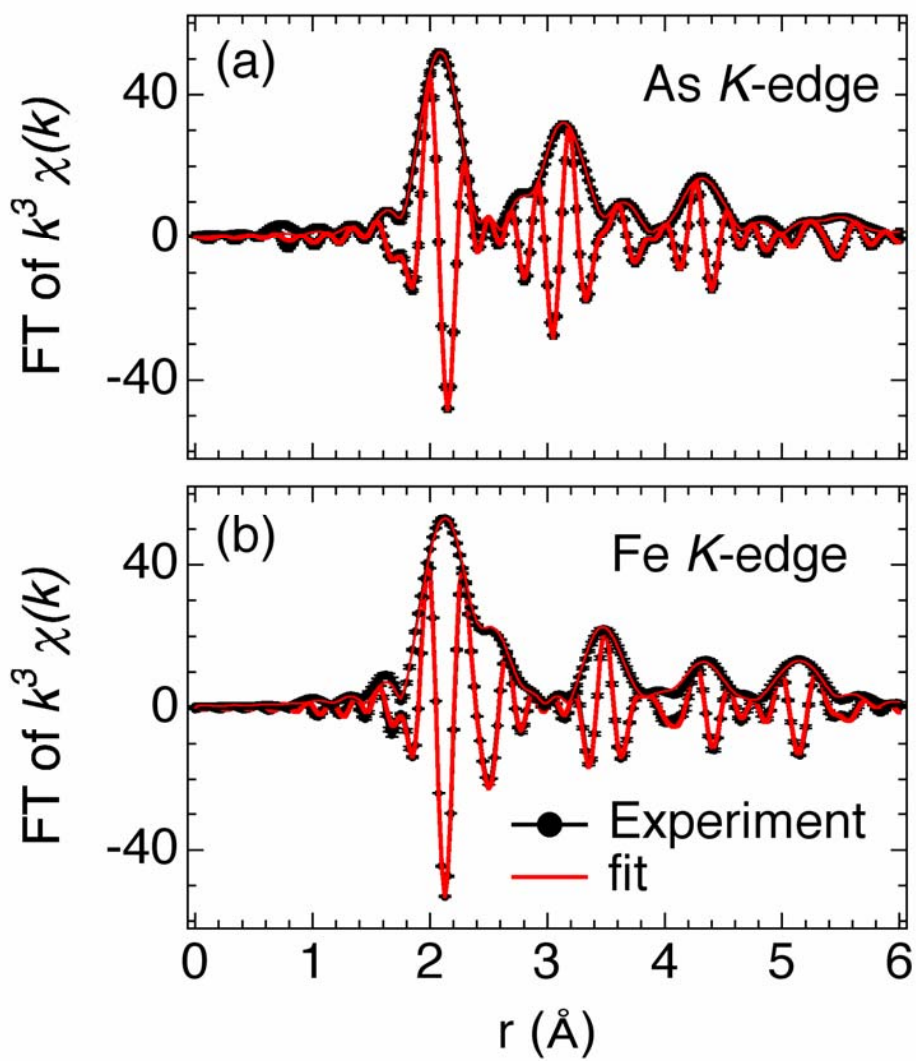

**Fig. 1** Fourier transform (FT) data of $CeFeAsO_{0.88}F_{0.12}$. (a) FT of As K edge data collected at 20 K together with the fits performed in r-space. (b) FT of Fe K edge data collected at 50 K together with the fits performed in r-space. FTs of As data are between 2.5 and 15.0 Å$^{-1}$, while FTs of Fe data are between 2.5 and 16.0 Å$^{-1}$, each using a Gaussian window 0.3 Å$^{-1}$ wide. Fits are between 1.2 and 6.1 Å.

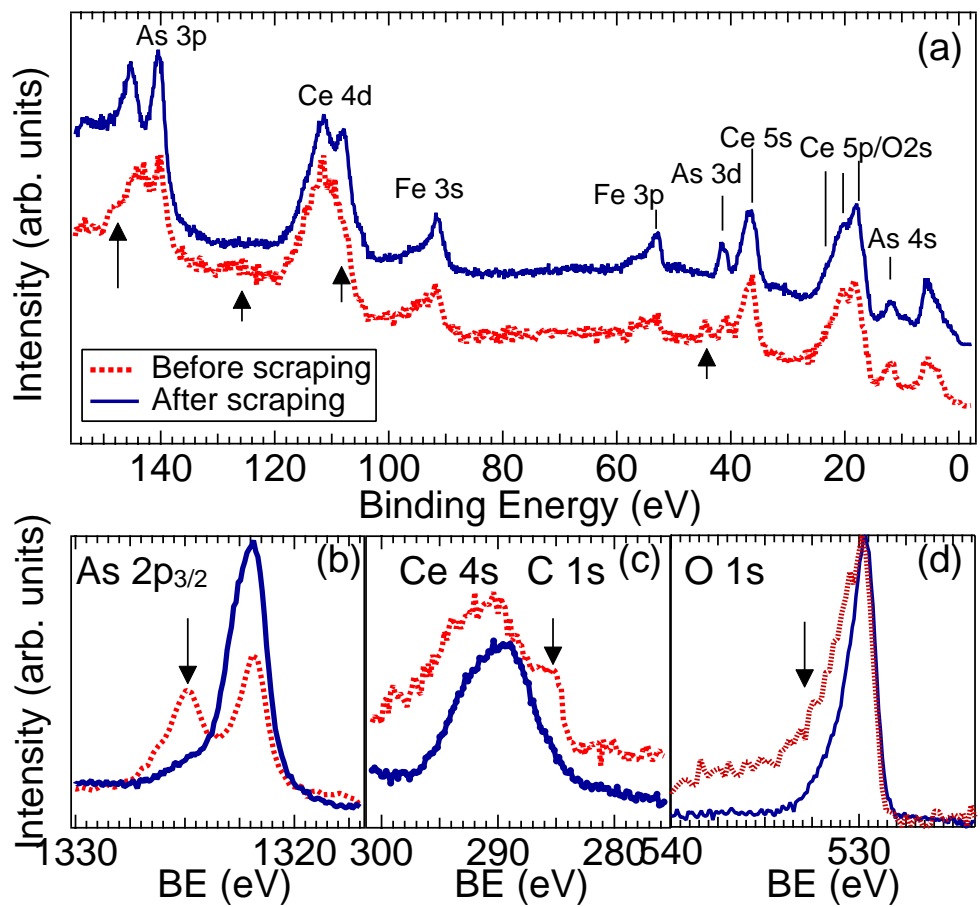

**Fig.2** HAXPES spectra from CeFeAsO$_{0.89}$F$_{0.11}$ measured with photon energy of 7596 eV at normal emission before and after scraping the surface in-situ with a diamond file. The signatures of surface contamination are indicated with arrows. (a) Overview spectrum of the shallow core levels; (b) As 2$p_{3/2}$ core level; (c) Ce 4$s$ core level with a C 1$s$ component visible before scraping; (d) O 1$s$ core level. In the clean sample, the C 1$s$ peak is not visible and the O 1$s$ peak has a single component with high line asymmetry.

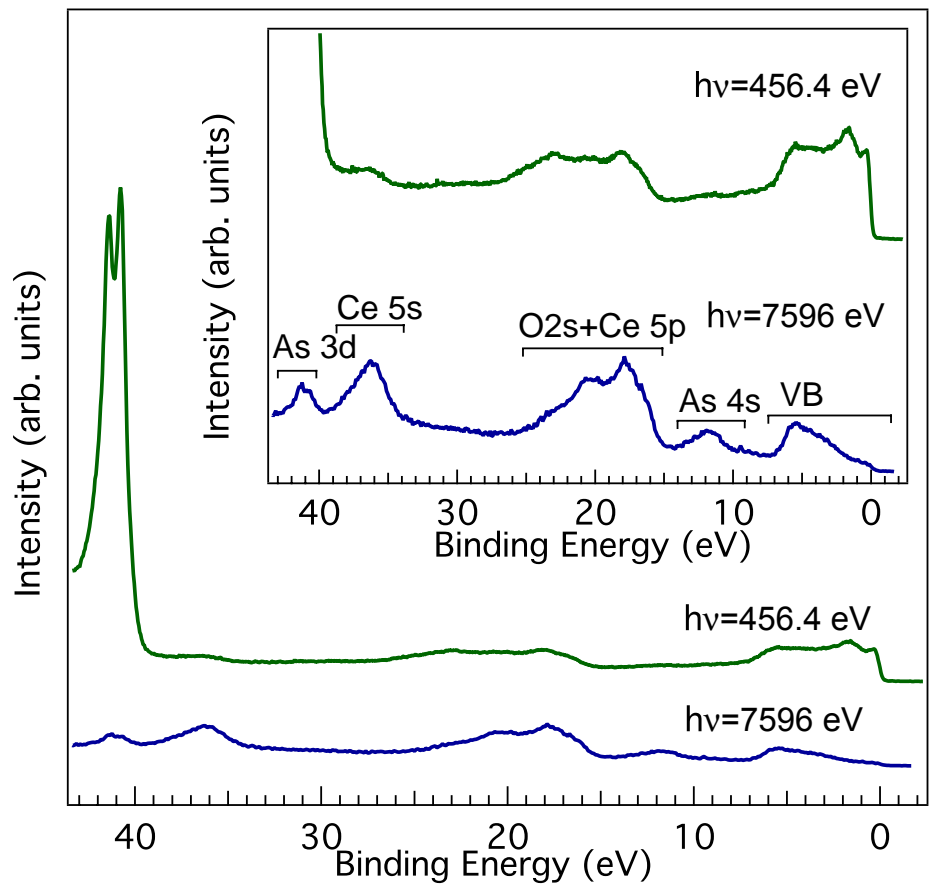

**Fig. 3** SXPES and HAXPES shallow core levels and valence band spectra of $CeFeAsO_{0.89}F_{0.11}$ measured at room temperature (297 K and 293 K) with photon energy of 456.4 eV and 7596 eV, respectively. The same spectrum is displayed in different intensity scales in order to show more clearly the shallow core levels close to the valence band. The clear changes in the relative intensity of the peaks are due to the higher cross section of *s* states with respect to *d* and *f* states in HAXPES with respect to SXPES.

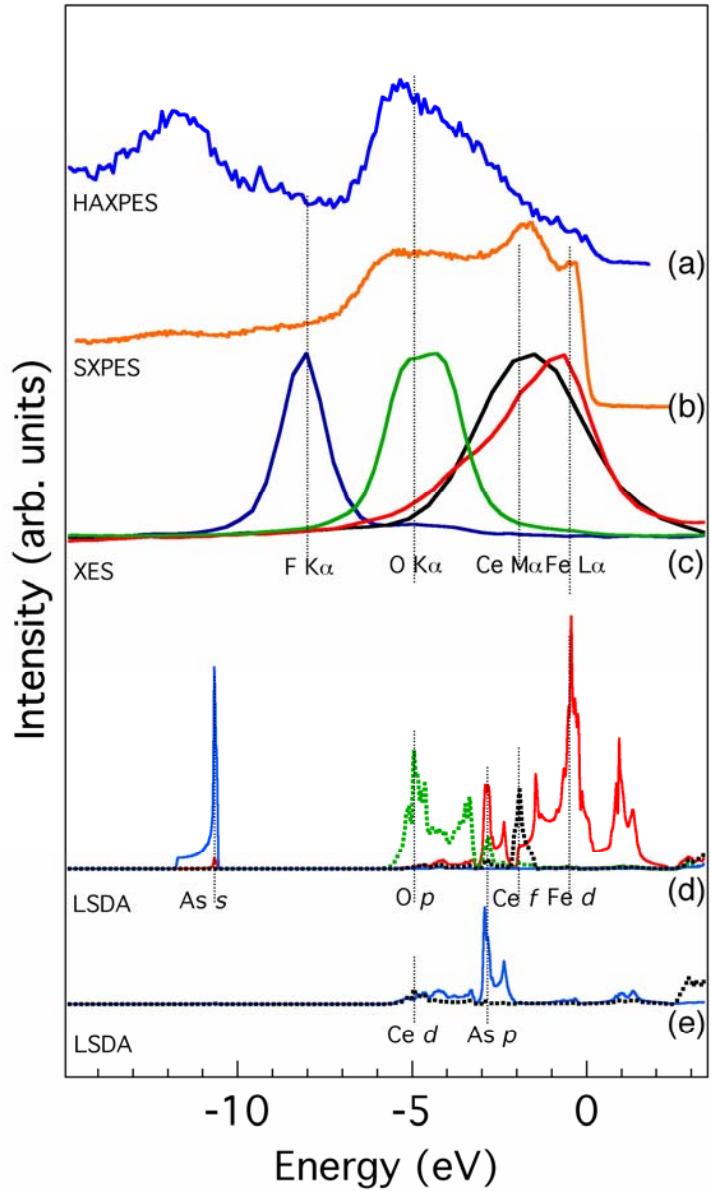

**Fig. 4** (a) Valence band HAXPES and (b) SXPES spectra measured with photon energy of 7596 eV and 175 eV, respectively, which provide a representation of the occupied total density of states weighted by the orbital cross section. (c) F $K_\alpha$, O $K_\alpha$, Ce $M_\alpha$ shallow core levels and Fe $L_\alpha$ XES spectra measured at room temperature. The F $K_a$, O $K_a$, Ce $M_\alpha$ and Fe$L_\alpha$ XES spectra have been measured with excitation energy of 696 eV, 534.1 eV, 882.4 eV, 717 eV, respectively. These near-threshold XES spectra are aligned to a common energy scale with respect to the core binding energies, reference to the Fermi level, enabling the decomposition of the valence band in the (F 2p, O 2p, Ce 4f and Fe 3d ) partial density of states components. (d) and (e) Partial As s, As p, O p, Ce f, Ce d Fe d DOS (average of majority and minority states) calculated for a virtual crystal with 10% doping.

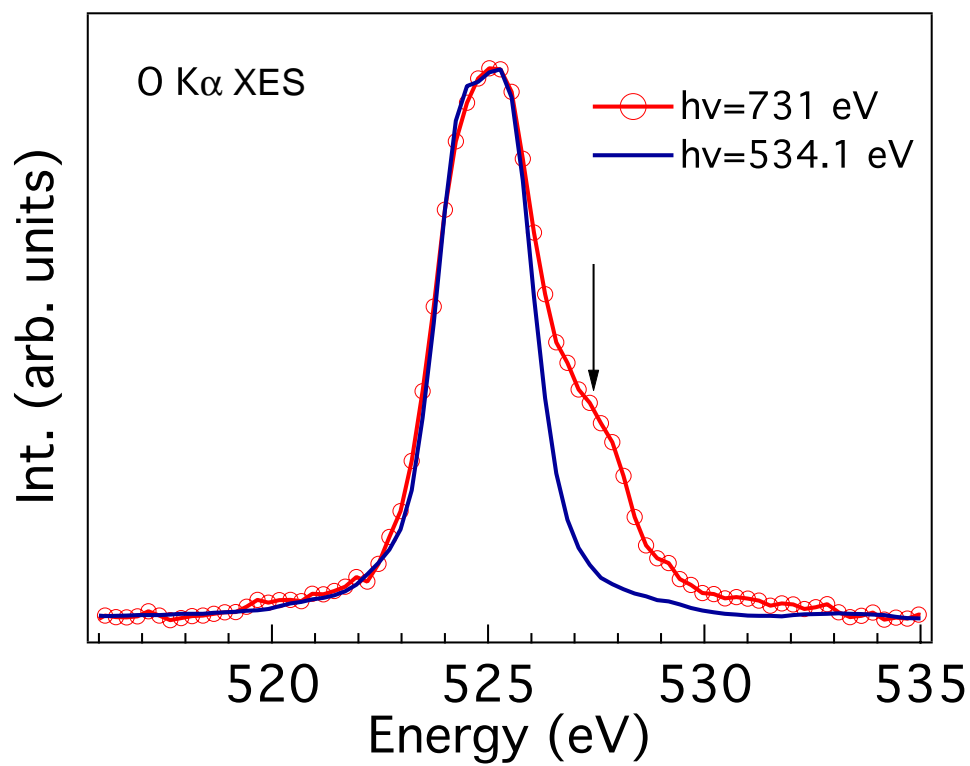

**Fig. 5** O $K_\alpha$ spectra measured with excitation energy of 534.1 eV and 731 eV. The high-energy shoulder appearing above the onset for $K^{-1}L^{-1}$ hole excitation is indicated with an arrow.

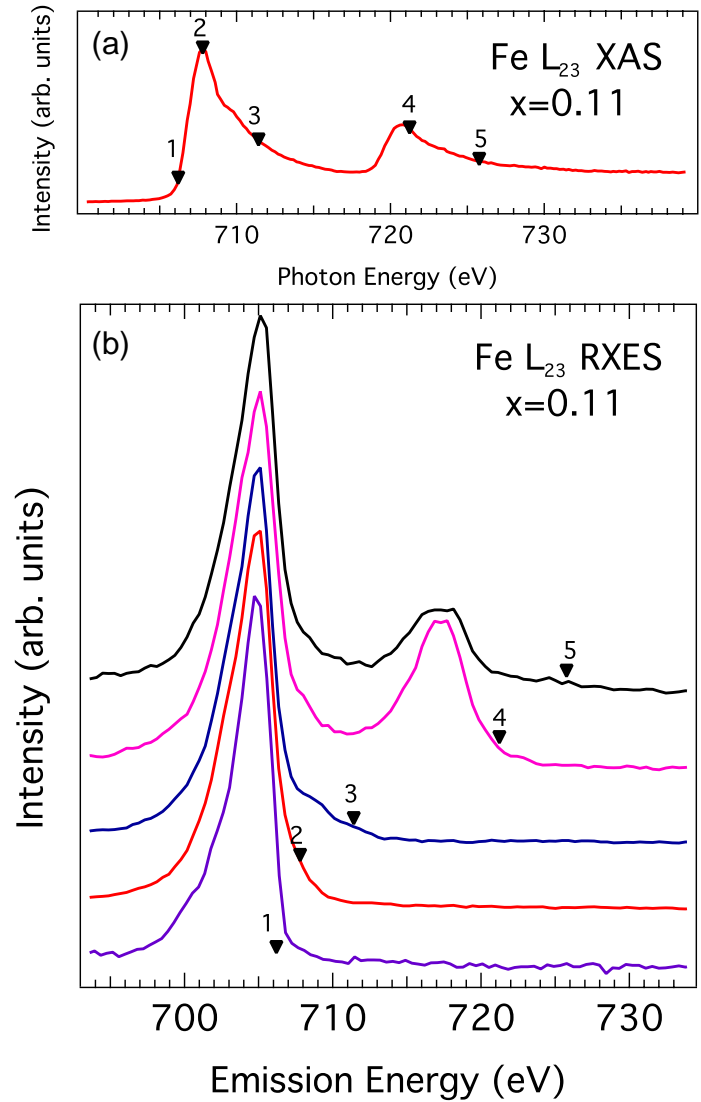

**Fig. 6** Fe $L_\alpha$ XES spectra (b) obtained at resonant excitation energies set across the Fe $L_{23}$ XAS edge (a). The resonant XES spectra display only structures at constant emission energy, at 705 eV and 717 eV. This behavior is typical of metallic systems such as Fe metal, consistent with an absence of strong correlations in this system.

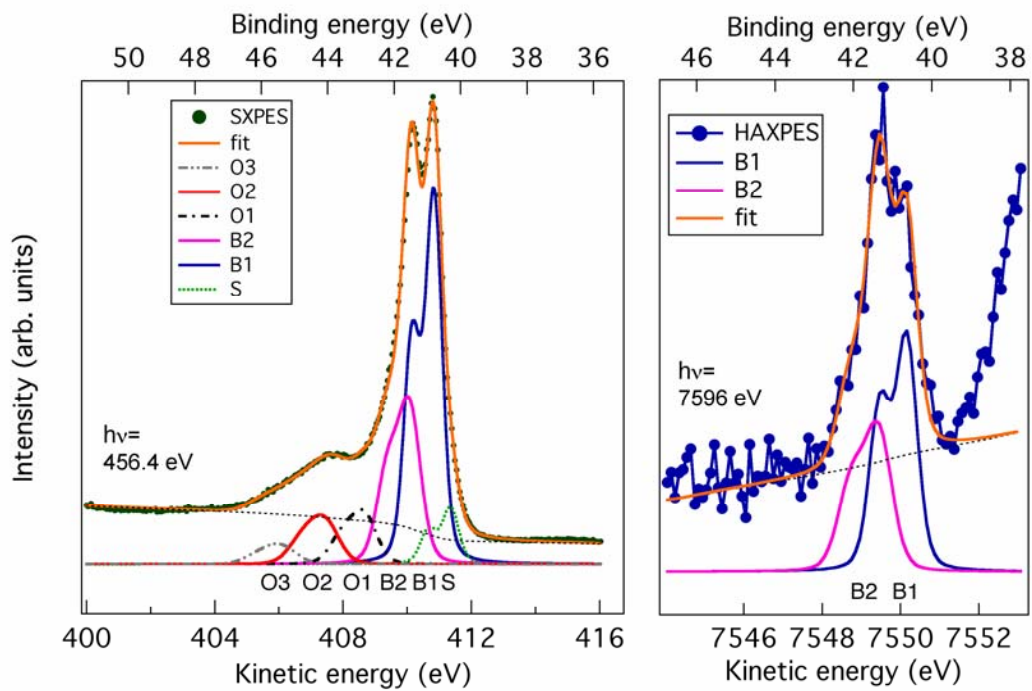

**Fig. 7** As 3$d$ core level spectra measured with two different photon energies (456.4 eV and 7596 eV) at room temperature and the curves resulting from the fitting procedure described in the text.

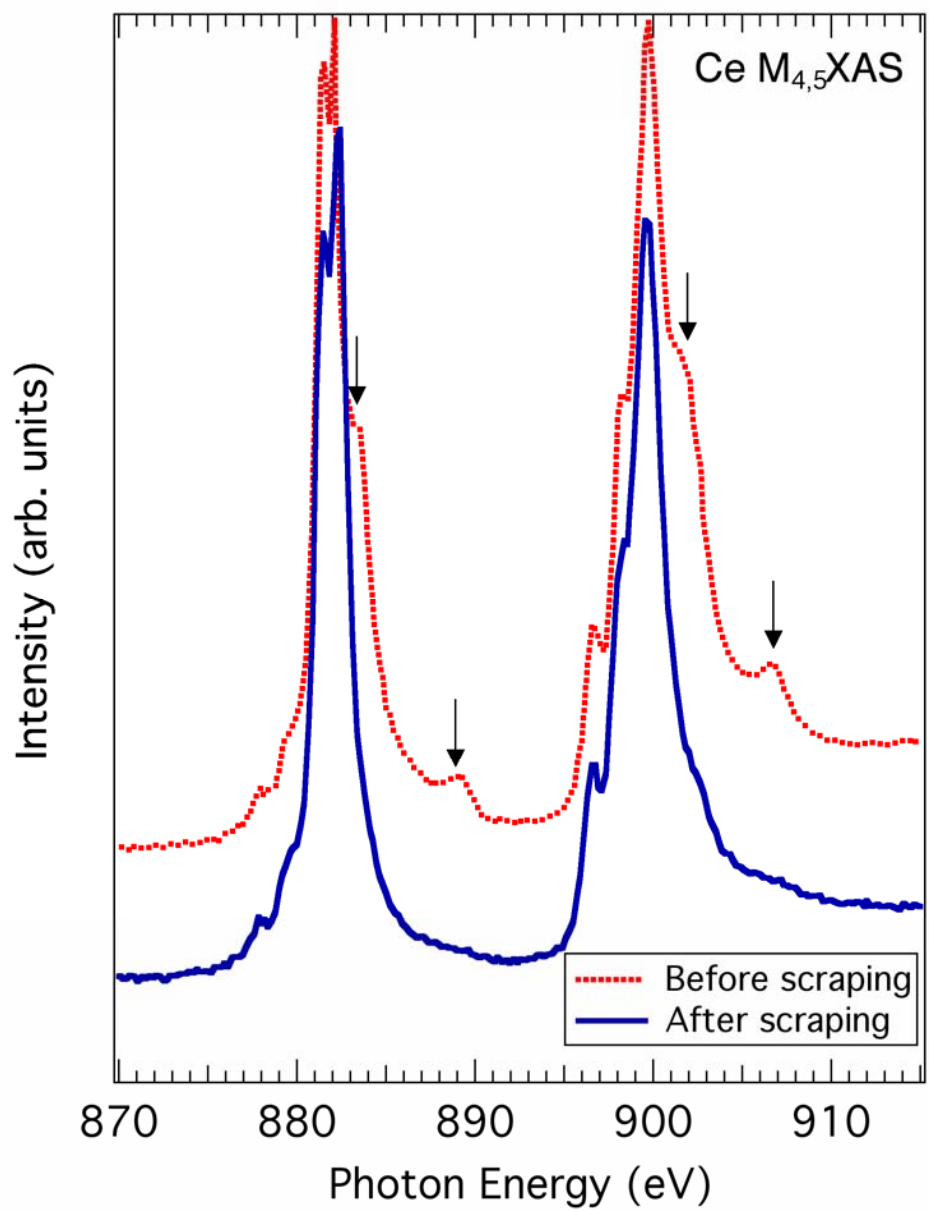

**Fig. 8** Ce $M_{4,5}$ XAS measured at room temperature from the CeFeAsO$_{0.89}$F$_{0.11}$ sample exposed to air and from the same sample after scraping in situ. The signatures of surface contamination are indicated with arrows.

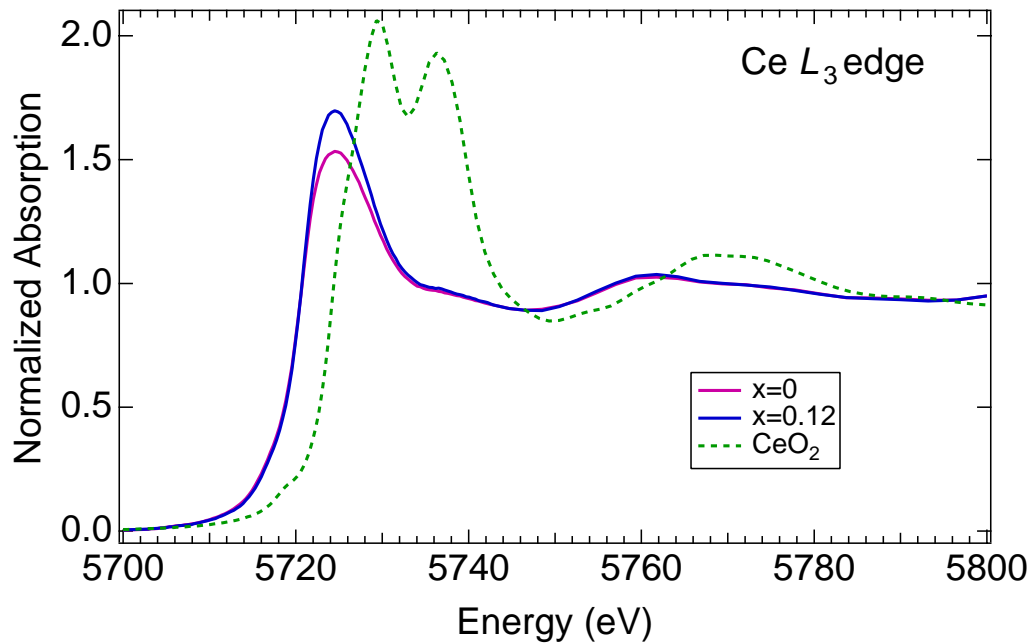

**Fig. 9.** Ce $L_3$-edge data at T=20 K from CeFeAsO$_{1-x}$F$_x$ (x=0, x=0.12). Data are calibrated by setting the energy of the first inflection point in the spectrum of a CeO$_2$ reference sample to 5724 eV measured at T=293 K.

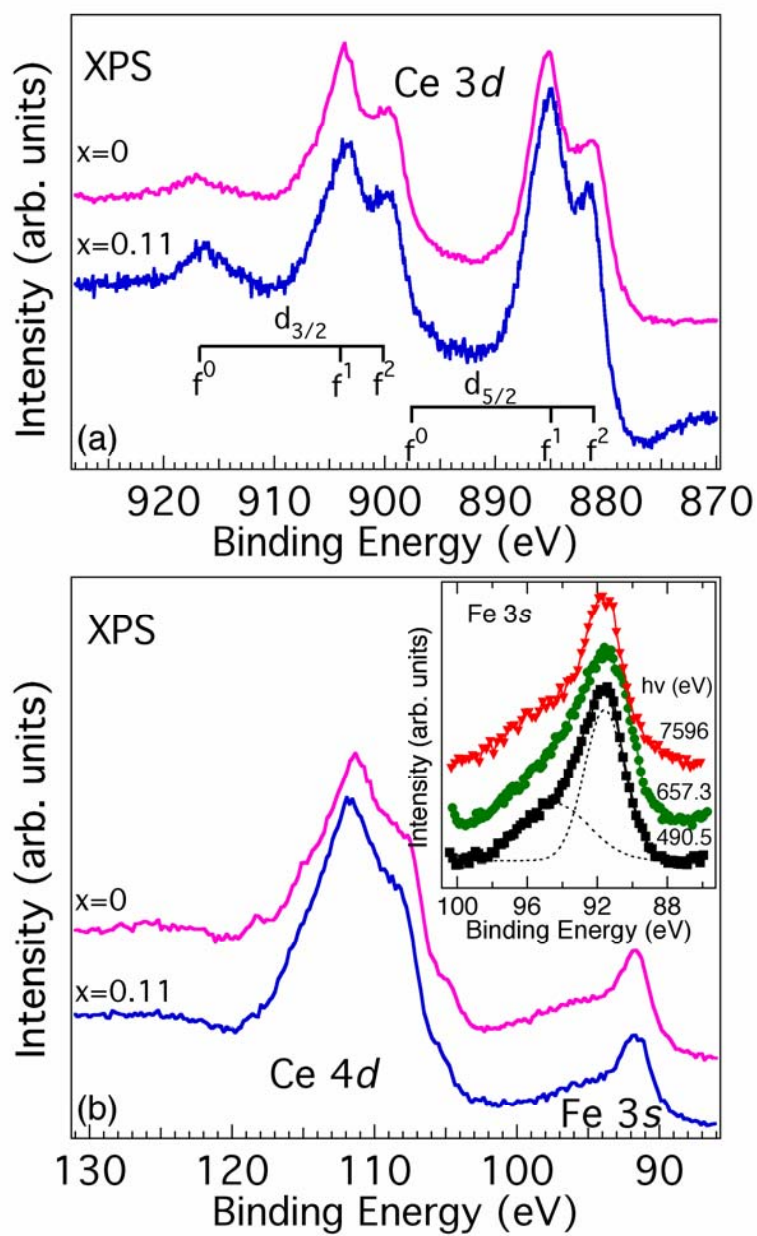

**Fig. 10** (a) Ce 3*d* core levels measured by HAXPES (hν=7596 eV) at room temperature from CeFeAsO$_{1-x}$F$_x$ (x=0, x=0.11). (b) Ce 4*d* and Fe 3*s* measured by HAXPES (hν=7596 eV) (curve B) and Fe 3*s* measured at room temperature by SXPES (hν=490.5 eV) (curve A) from CeFeAsO$_{1-x}$F$_x$ (x=0, x=0.11). The inset shows the Fe 3*s* measured with HAXPES and SXPES from CeFeAsO$_{0.89}$F$_{0.11}$.

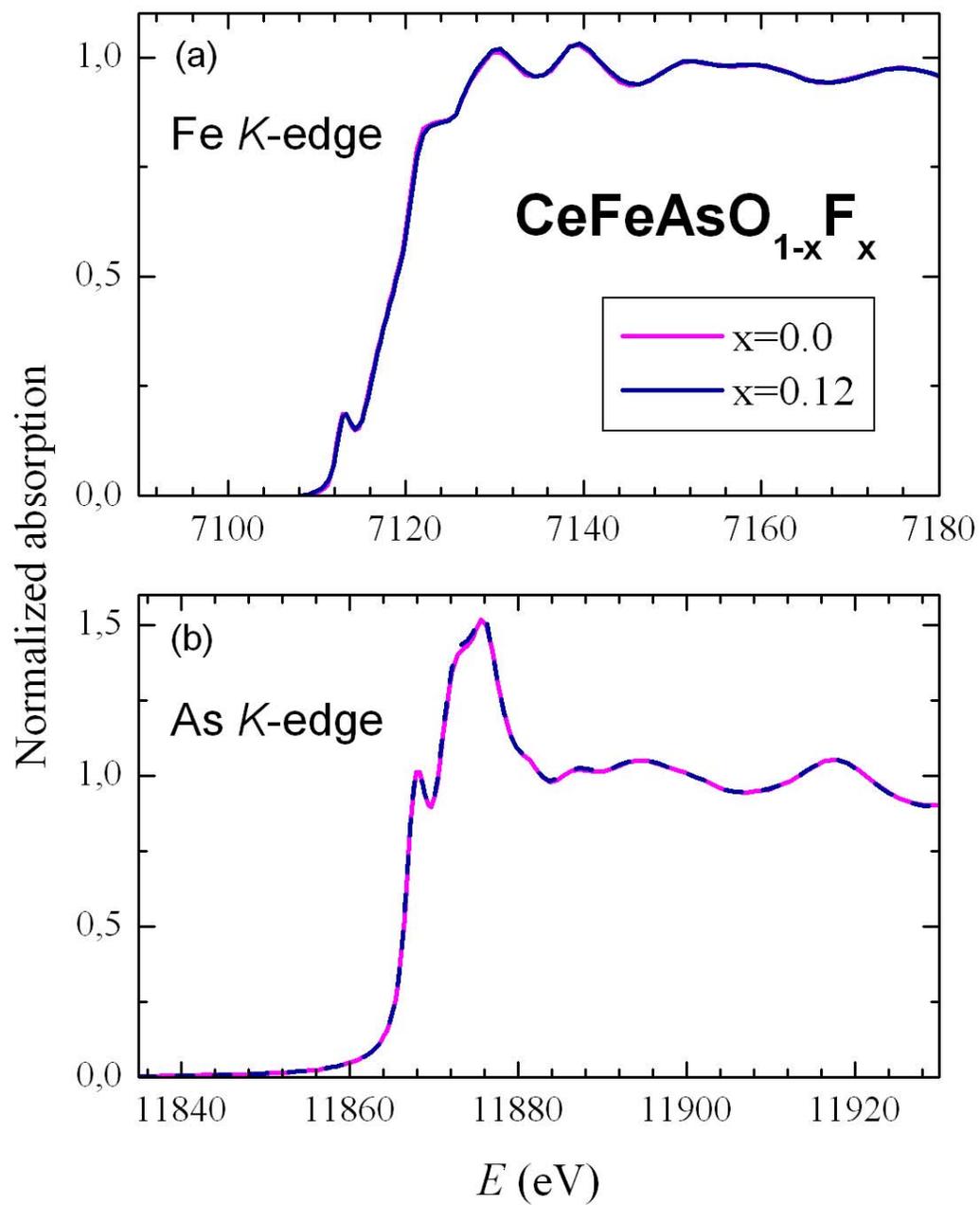

**Fig. 11** (a) Fe *K*-edge and (b) As *K*-edge XAS data at T=20 K from CeFeAsO$_{1-x}$F$_x$ (x=0, x=0.12).

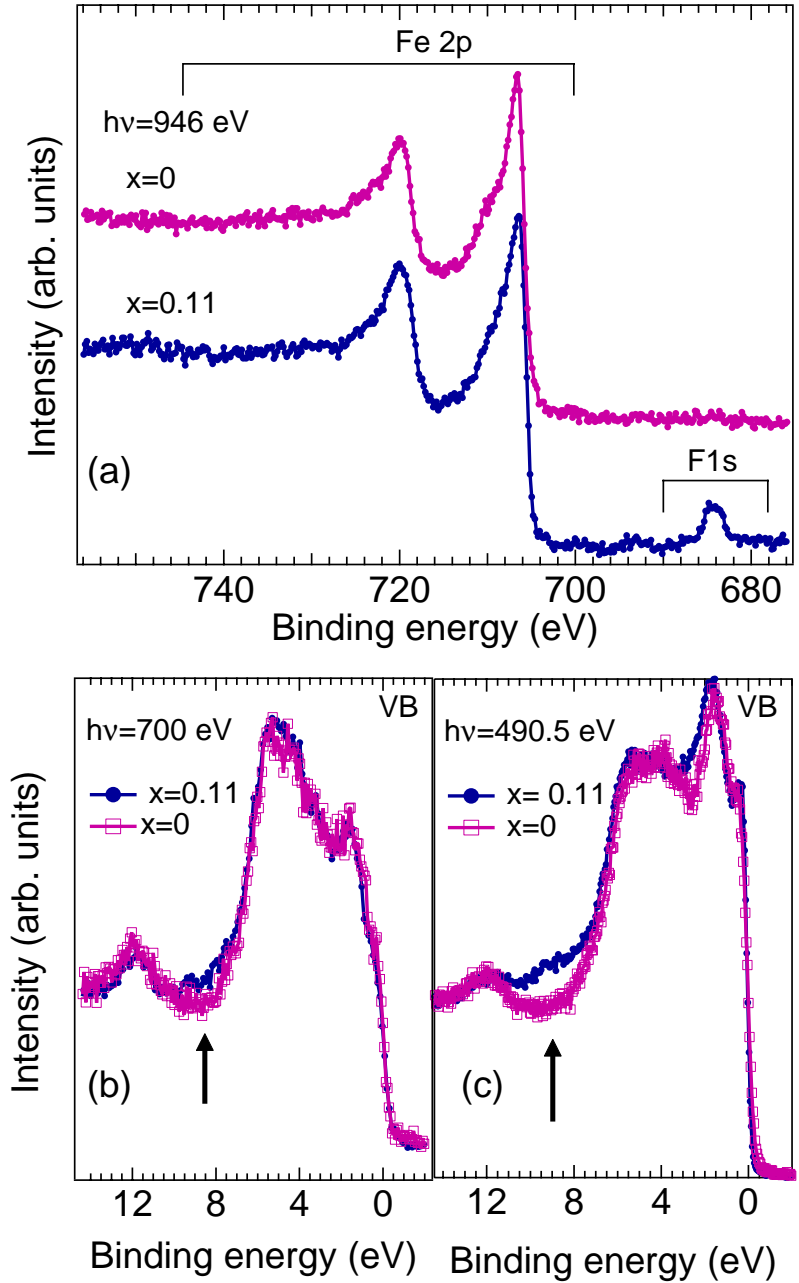

**Fig. 12** Comparison of PES spectra taken at T=297 K from CeFeAsO$_{1-x}$F$_x$ (x=0, x=0.11). (a) Fe 2p XPS measured with hν=946 eV at room temperature. (b) Valence band measured with hν=700 eV. (c) Valence band measured with hν=490.5 eV. Some differences in the valence band (marked with an arrow) are observed at binding ~ 9 eV, likely due to the F 2p spectral weight in the x=0.11 sample.

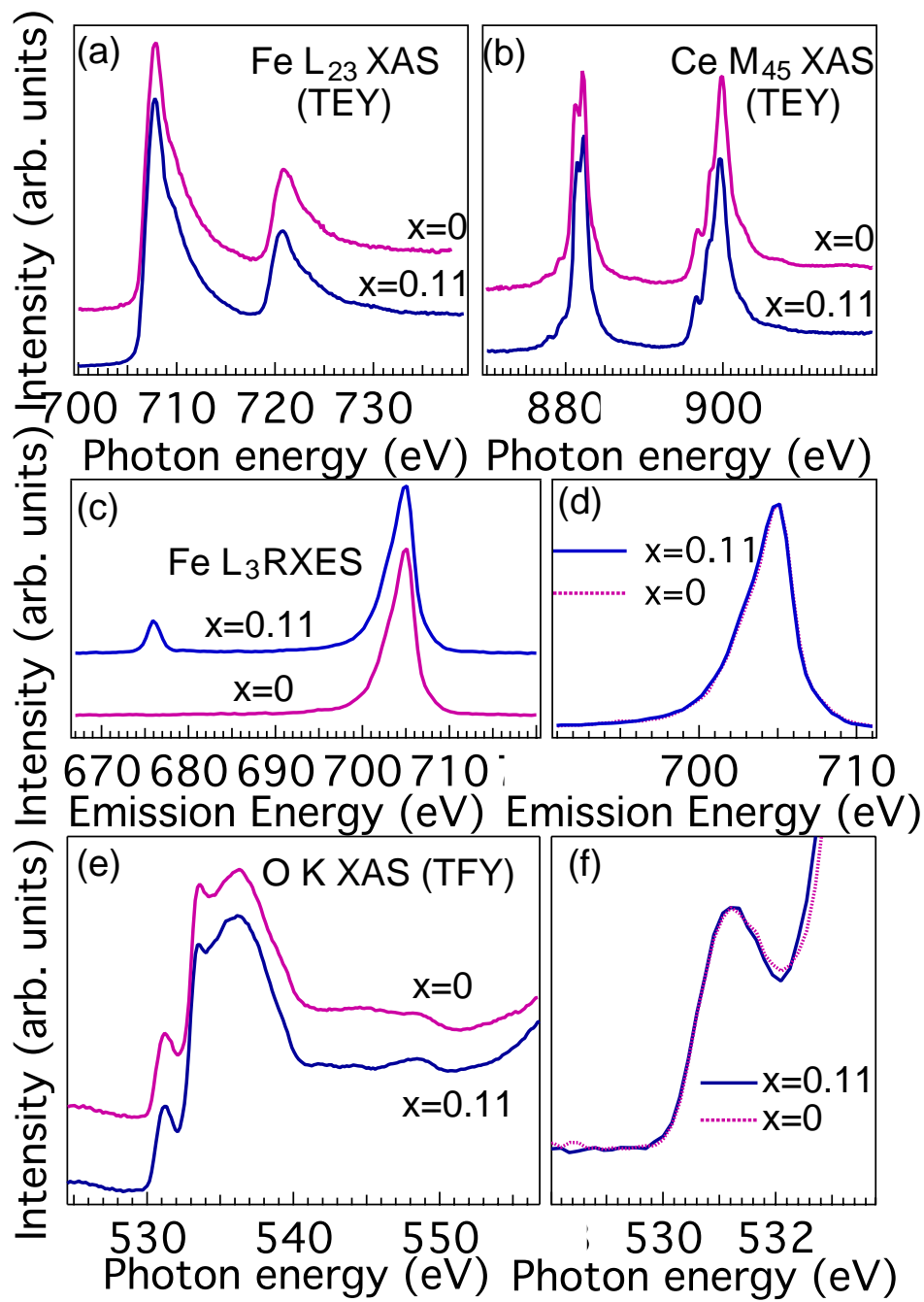

**Fig. 13** Comparison of XAS and XES spectra of CeFeAsO$_{1-x}$F$_x$ (x=0, x=0.11). (a) Fe $L_{23}$ XAS; (b) Ce $M_{45}$ XAS. (c) Fe $L_\alpha$ and F $K_\alpha$ XES (d) F $L_\alpha$ XES (e) O $K$ XAS normalized to the prepeak, measured in total fluorescence yield with the sample biased at +600 V (f) Enlarged view of the prepeak region of the O $K$ XAS (showing no energy shift between the spectra). No clear doping dependence is found in the spectra.

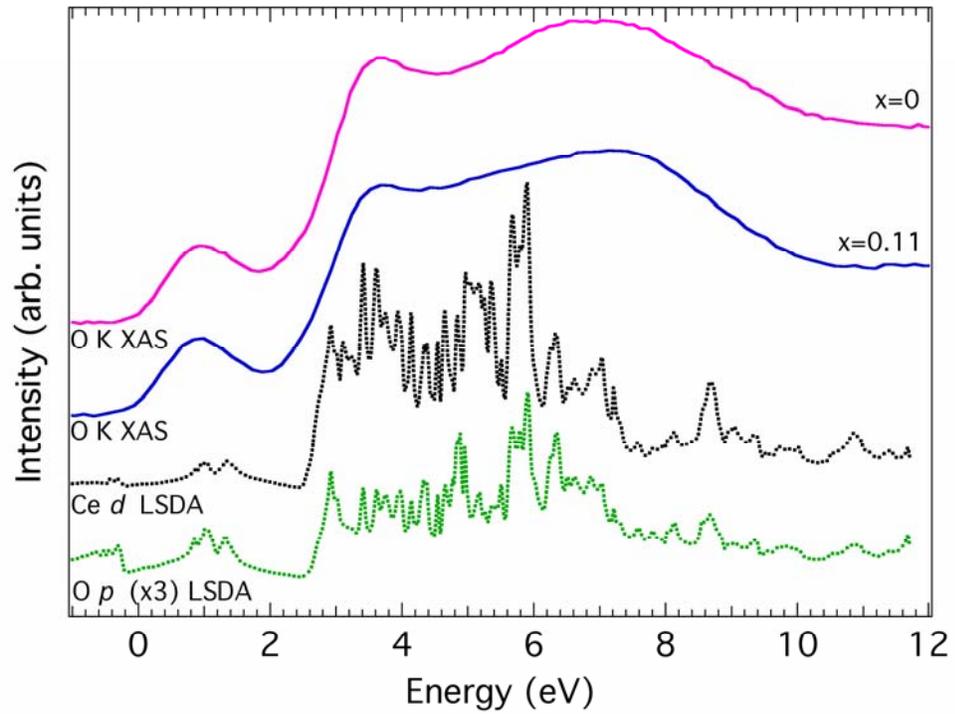

**Fig 14** O *K* XAS measured in total electron yield from CeFeAsO$_{1-x}$F$_x$ (x=0, x=0.11) compared to partial O *p* and Ce *d* DOS calculated using the LSDA approach. The energy scale of the O *K* XAS spectrum has been arbitrarily offset by -525.5 eV in order to compare the experimental data with the calculations on the same scale.